\documentclass[12pt]{article}
\usepackage{graphics, color}
\usepackage{graphicx}
\usepackage{amssymb}

\newcommand{\sect}[1]{\section{#1}\setcounter{equation}{0}}

\begin{document}


\begin{titlepage}
\bigskip
\bigskip\bigskip\bigskip
\centerline{\Large \bf Analytic Study of Small Scale Structure on Cosmic Strings}
\bigskip\bigskip\bigskip

\centerline{\large Joseph Polchinski}
\bigskip
\centerline{\em Kavli Institute for Theoretical Physics}
\centerline{\em University of California}
\centerline{\em Santa Barbara, CA 93106-4030} \centerline{\tt joep@kitp.ucsb.edu}
\bigskip
\bigskip
\centerline{\large Jorge V. Rocha}
\bigskip
\centerline{\em Department of Physics}
\centerline{\em University of California}
\centerline{\em Santa Barbara, CA 93106} \centerline{\tt jrocha@physics.ucsb.edu}
\bigskip
\bigskip
\bigskip\bigskip


\begin{abstract}
The properties of string networks at scales well below the horizon are poorly understood, but they enter critically into many observables.  We argue that in some regimes, stretching will be the only relevant process governing the evolution.  
In this case, the string two-point function is determined up to normalization: the fractal dimension approaches one at short distance, but the rate of approach is characterized by an exponent that plays an essential role in network properties.
The smoothness at short distance implies, for example, that cosmic string lensing images are little distorted.  We then add in loop production as a perturbation and find that it diverges at small scales.  This need not invalidate the stretching model, since the loop production occurs in localized regions, but it implies a complicated fragmentation process.  Our ability to model this process is limited, but we argue that loop production peaks a few orders of magnitude below the horizon scale, without the inclusion of gravitational radiation.  We find agreement with some features of simulations, and interesting discrepancies that must be resolved by future work.
\end{abstract}
\end{titlepage}
\baselineskip = 16pt

\sect{Introduction}

The evolution of cosmic string networks is a challenging problem.  The need to consider large ratios of length and time scales makes a complete numerical analysis impossible, while the nonlinearity of the process defeats a purely analytic treatment.   Thus a full understanding of most of the signatures of cosmic strings will require a careful combination of analytic and numerical approaches.  This is needed both to establish precisely the current bounds on the dimensionless cosmic string tension $G\mu$, and to anticipate what will be the most sensitive future measurements.  Also, if cosmic strings are one day discovered, a precise understanding of the network properties will be needed in order to distinguish different microscopic models.

On scales close to the horizon size, the networks are reasonably well understood from simulations~\cite{Albrecht:1989mk,Bennett:1989yp,Allen:1990tv}.\footnote{For reviews of network properties and other aspects of cosmic strings see Refs.~\cite{VilShell,Hindmarsh:1994re}.}
In particular, there are a few dozen long strings crossing any horizon volume.  On shorter scales, however, the situation is far less clear.  There is a nontrivial short distance structure on the strings, which arises because the intercommutation process produces kinks~\cite{Bennett:1987vf}.  There have been many previous analytic and numerical studies of this, but there is no general consensus as to its nature.

In particular, the size at which typical loops form is uncertain to tens of orders of magnitude.  One widely-held assumption has been that gravitational radiation is necessary in order for cosmic string networks to scale~\cite{ACK1993}, and that it determines the size of loops.  If so, the loops will be parametrically smaller than the horizon scale, by a power of $G\mu$~\cite{Siemens:2002dj} (even shorter scales have been suggested~\cite{Vincent:1996rb}).  More recent work appears to be converging on loop sizes at a fixed fraction of the horizon scale, but even here there are significant differences.  Ref.~\cite{Vanchurin:2005pa} suggests that loops form at around $0.1$ times the horizon scale, whereas Refs.~\cite{RSB2005,MarShell2005} (and some portion of the conventional wisdom) suggest a scale several orders of magnitude smaller.  The larger loops would lead to enhanced signatures of several types, and tighter bounds on $G\mu$~\cite{Olum:2006at}.

In a system with a large hierarchy of scales, one might hope that analytic methods would be particularly useful in separating the processes occurring at different scales,
while numerical methods would be needed to understand the nonlinearities at a given scale.  This is the philosophy that we will attempt to implement.  We focus on a microscopic description, similar in spirit to Ref.~\cite{Embacher:1992zk} and in particular Ref.~\cite{ACK1993}.  An important difference from Ref.~\cite{ACK1993} is that we are less ambitious: that work was largely directed at understanding the horizon-scale structure quantitatively, whereas we are only interested in shorter scales.  Also, we do not attempt to reduce the dynamics to a parameterized model but rather focus on the full two-point function; this two-point function appears to be characterized by a critical exponent, which did not enter into previous work.  Finally, our (tentative) conclusions are opposite to Ref.~\cite{ACK1993}, in that we believe that the network scales even without gravitational radiation.

The full microscopic equations for the string network~\cite{Embacher:1992zk,ACK1993} appear to be too complicated to solve, and so we model what we hope are the essential physical processes.  We have had some success in matching results from simulations, but there are also some discrepancies which may indicate additional processes that must be included.

In section~2 we consider the evolution of a small segment on a long string.  If the rate of string intercommutations is fixed in horizon units, then over a range of scales the only important effect would be stretching due to the expansion of the universe.  We are then able to determine the two-point functions characterizing the small scale structure.  We find that the string is actually rather smooth, in agreement with recent simulations~\cite{MarShell2005}: its fractal dimension approaches one as we go to smaller scales.  There is a nontrivial power law, but this appears in the {\it approach} of the fractal dimension to one; that is, the critical exponent $\chi$ is related to the power spectum of perturbations on the long string.  The two-point functions depend on two parameters that must be taken from simulations.  One of these is the mean $v^2$ of an element of the long string; this determines the exponent $\chi$.  The second parameter is the normalization of the two-point function.  Our results for the two-point function match reasonably well with the simulations~\cite{MarShell2005} over a range of scales, but there is substantial deviation at the shortest scales.  It remains to be seen whether this is due to transient effects in the simulations, or real effects that we have omitted.

In section~3 we study the effects of the small scale structure on string lensing.  Because the string is fairly smooth, these effects are not as dramatic as has been considered in some previous work.  However, there are calculable deviations from perfect lensing.  We also consider, for lensing by a long string, the likely trajectory of the string relative to the axis of a given lensed object.   We discuss some deviations from gaussianity in the small scale structure, due to kinks.

In section~4 we add in loop production as a perturbation.  Even though the strings are relatively smooth at short distance, we find that the total rate of loop production diverges for small loops;  the rate of divergence appears to agree with recent simulations~\cite{RSB2005}.
This divergence does not invalidate our stretching model, because the loop production is localized to regions where the left- and right-moving tangents ${\bf p}_{\pm}$ are approximately equal, but it points to a complicated fragmentation process, cascading to smaller and smaller loops.  We are not able to follow the fragmentation process analytically at present, but we give an analytic argument as to why the cascade should terminate, leaving loops at some small but fixed multiple of the horizon scale.  We also point out an interesting puzzle related to loop velocities, which again points to a complicated fragmentation process.

In section~5 we discuss various applications and future directions.  The behavior of string networks is notoriously complicated, and it appears to remain so even when one focusses on small scales.  Thus we view our work as part of an ongoing dialogue between analytic calculations and simulations, which we hope will lead to a more complete and precise picture. 

\vspace{2pc}

\sect{The stretching regime}

\subsection{Assumptions}

We consider ``vanilla'' cosmic strings, a single species of local string without superconducting or other extra internal degrees of freedom.  The evolution of a network of such strings is dictated by three distinct processes.  First, the expansion of the universe stretches the strings: on scales larger than the horizon scale $d_H$ irregularities on a string are just conformally amplified, but on smaller scales the string effectively straightens~\cite{Vilenkin1981}.  This is described by the Nambu action in curved spacetime.  Second, gravitational radiation also has the effect of straightening the strings, but it is significant only below a length scale proportional to $d_H$ and to a positive power of the dimensionless string coupling $G\mu$~\cite{BB1989,Sakellariadou:1990ne,Hindmarsh:1990xi,Quashnock:1990wv,Siemens:2002dj}.  Since this is parametrically small at small $G\mu$, we will ignore this effect for the present purposes.  Finally, intercommutations play an important role in reaching a scaling solution, in particular through the formation of closed loops of string.  At first we shall neglect this effect but we will be forced to return to this issue in section~4.

Let us consider the evolution of a small segment on a long string.  We take the segment to be very short compared to the horizon scale, but long compared to the scale at which gravitational radiation is relevant.  The scaling property of the network implies that the probability per Hubble time for this segment to be involved in a long string intercommutation event is proportional to its length divided by $d_H$, and so for short segments the intercommutation rate per Hubble time will be small.  Formation of a loop much larger than the segment might remove the entire segment from the long string, but this should have little correlation with the configuration of the segment itself, and so will not affect the probability distribution for the ensemble of short segments.  Formation of loops at the size of the segment and smaller could affect this distribution, and the results of section~4 will indicate that the production of small loops is large, but this process takes place only in localized regions where the left- and right-moving tangents are approximately equal.  Thus, there is a regime where stretching is the only relevant process.

If we follow a segment forward in time, its length increases but certainly does so more slowly than the horizon scale $d_H$, which is proportional to the FRW time $t$.  Thus the length divided by $d_H$ decreases, and therefore so does the rate of intercommutation.  If we follow the segment backward in time, its length eventually begins to approach the horizon scale, and the probability becomes large that we encounter an intercommutation event.  Our strategy is therefore clear.  For the highly nonlinear processes near the horizon scale we must trust simulations.  At a somewhat lower scale we can read off the various correlators describing the behavior of the string, and then evolve them forward in time using the Nambu action until we reach the gravitational radiation scale.  The small probability of an intercommutation involving the short segment can be added as a perturbation.  This approach is in the spirit of the renormalization group, though with long and short distances reversed.

\subsection{Two-point functions at short distance}

In an FRW spacetime,
\begin{eqnarray}
ds^2 &=& -dt^2 + a(t)^2 d{\bf x} \cdot d{\bf x} \nonumber\\
&=& a(\tau)^2 (-d\tau^2 + d{\bf x} \cdot d{\bf x})\ ,
\end{eqnarray}
the equation of motion governing the evolution of a cosmic string is~\cite{TurokBhat1984}
\begin{equation}
\ddot{\bf x} + 2\,\frac{\dot{a}}{a}\,(1-\dot{\bf x}^2)\, \dot{\bf x} = \frac{1}{\epsilon} \left( \frac{{\bf x}'}{\epsilon} \right)' \ .
\label{EOM}
\end{equation}
Here $\epsilon$ is given by
\begin{equation}
\epsilon \equiv \left( \frac{{\bf x}'^2}{1-\dot{\bf x}^2} \right)^{1/2}.
\label{epsilon}
\end{equation}
These equations hold in the transverse gauge, where $\dot{\bf x}\cdot{\bf x}'=0$. Dots and primes refer to derivatives relative to the conformal time $\tau$ and the spatial parameter $\sigma$ along the string, respectively. The evolution of the parameter $\epsilon$ follows from equation (\ref{EOM}),
\begin{equation}
\frac{\dot{\epsilon}}{\epsilon} = -2\,\frac{\dot{a}}{a} \, \dot{\bf x}^2.
\label{epsev}
\end{equation}

From the second derivative terms it follows that signals on the string propagate to the right and left with $d\sigma = \pm d\tau/\epsilon$.  Thus the structure on a short piece of string at a given time is a superposition of left- and right-moving segments, and it is these that we follow in time.   In an expanding spacetime the left- and right-moving waves interact --- they are not free as in flat spacetime.

From Eq.~(\ref{epsev}) it follows that the time scale of variation of $\epsilon$ is the Hubble time, and so to good approximation we can replace $\dot{\bf x}^2$ with the time-averaged $\bar v^2$
(bars will always refer to RMS averages), giving $\epsilon \propto a^{-2\bar v^2}$ as a function of time only.\footnote{The transverse gauge choice leaves a gauge freedom of time-independent $\sigma$ reparameterizations.  A convenient choice is to take $\epsilon$ to be independent of $\sigma$ at the final time, and then $\epsilon$ will be $\sigma$-independent to good approximation on any horizon length scale in the past.}
  From the definition of $\epsilon$ it then follows that the energy of a segment of string of coordinate length $\sigma$ is
$E = \mu a(\tau) \epsilon(\tau) \sigma$.
For simplicity we will refer to $E/\mu$ as the length $l$ of a segment,
\begin{equation}
l = a(\tau) \epsilon(\tau) \sigma\ ,
\end{equation}
though this is literally true only in the rest frame.  The scale factor is
\begin{equation}
a \propto t^\nu \propto \tau^{\nu'}\ ,\quad \nu' = \nu/(1 - \nu)\ ,
\end{equation}
where
\begin{eqnarray}
\mbox{radiation domination:}&& \nu = 1/2\ ,\quad \nu' = 1\ ,\quad \bar v^2 \approx 0.41 \ ,\nonumber\\
\mbox{matter domination:}&& \nu = 2/3\ ,\quad \nu' = 2\ ,\quad \bar v^2 \approx 0.35 \ .
\end{eqnarray}
The mean RMS velocities for points on long strings are taken from simulations~\cite{MarShell2005}. It follows that
\begin{equation}
l \propto \tau^{\zeta'} \propto t^{\zeta}\ , \quad \zeta' = (1 - 2\bar v^2) \nu' \ , \quad
\zeta = (1 - 2\bar v^2) \nu \ . \label{stretch}
\end{equation}
In the radiation era, $\zeta_r \sim 0.1$ and in the matter era $\zeta_m \sim 0.2$.  Thus the physical length of the segment grows in time, but more slowly than the comoving length~\cite{Vilenkin1981}, and much more slowly than the horizon length $d_H \propto t$.

For illustration, consider a segment of length $(10^{-6}\; {\rm to}\; 10^{-7}) d_H$, as would be relevant for lensing at a separation of a few seconds and a redshift of the order of $z\simeq 0.1$.  According to the discussion above, $l / d_H$ depends on time as $t^{\zeta - 1} \sim t^{-0.8}$ in the matter era.  Thus the length of the segment would have been around a hundredth of the horizon scale at the radiation-to-matter transition.  In other words, it is the nonlinear horizon scale dynamics in the radiation epoch that produces the short-distance structure that is relevant for lensing today, in this model.  This makes clear the limitation of simulations by themselves for studying the small scale structure on strings, as they are restricted to much smaller dynamical ranges.

In terms of left- and right-moving unit vectors $\bf{p}_\pm \equiv \dot{\bf{x}} \pm \frac{1}{\epsilon}\bf{x}'$, the equation of motion~(\ref{EOM}) can be written as~\cite{Albrecht:1989mk}
\begin{equation}
\dot{\bf p}_\pm \mp \frac{1}{\epsilon} {\bf p}'_\pm = - \frac{\dot{a}}{a} \left[ {\bf p}_\mp - ({\bf p}_+ \cdot {\bf p}_-)\, {\bf p}_\pm \right].
\label{EOM2}
\end{equation}
We will study the time evolution of the left-moving product ${\bf p}_+(\tau,\sigma) \cdot {\bf p}_+(\tau,\sigma')$.  For this it is useful to change variables from $(\tau,\sigma)$ to $(\tau,s)$ where $s$ is constant along the left-moving characteristics,
$\dot s - s'/\epsilon = 0$.
Then
\begin{eqnarray}
\partial_\tau ({\bf p}_+ (s,\tau) \cdot {\bf p}_+(s',\tau)) &=&
 - \frac{\dot a}{a} \biggl( {\bf p}_-(s,\tau)\cdot{\bf p}_+(s',\tau)+  {\bf p}_+(s,\tau)\cdot{\bf p}_-(s',\tau) \nonumber\\
&&\hspace{-60pt} +  \alpha (s,\tau)\,{\bf p}_+(s,\tau)\cdot{\bf p}_+(s',\tau) + \alpha  (s',\tau)\,{\bf p}_+(s,\tau)\cdot{\bf p}_+(s',\tau) \biggr)\ , \label{bilinear}
\end{eqnarray}
where $\alpha = - {\bf p}_+ \cdot{\bf p}_- = 1 - 2v^2$.

The equations of motion~(\ref{EOM2}, \ref{bilinear}) are nonlinear and do not admit an analytic solution, but they simplify when we focus on the small scale structure.  If ${\bf p}_+ (s,\tau)$ were a smooth function on the unit sphere, we would have $1 - {\bf p}_+ (s,\tau) \cdot {\bf p}_+(s',\tau) = O([s-s']^2)$ as $s'$ approaches $s$.  We are interested in any structure that is less smooth than this, meaning that it goes to zero more slowly than $[s-s']^2$.  For this purpose we can drop any term of order $[s-s']^2$ or higher in the equation of motion (smooth terms of order $s-s'$ cancel because the function is even).

Consider the product ${\bf p}_-(s,\tau) \cdot {\bf p}_+(s',\tau)$.  The right-moving characteristic through $(s,\tau)$ and the left-moving characteristic through $(s',\tau)$ meet at a point $(s,\tau-\delta)$ where $\delta$ is of order $s - s'$.\footnote
{Explicitly, for given $\tau$ we could choose coordinates where $\epsilon(\tau) = 1$ and $s(\tau',\sigma) = \sigma + \tau' - \tau+ O( [\tau' - \tau]^2)$, and then $\delta = (s-s') /2$.}  Eq.~(\ref{EOM2}) states that ${\bf p}_+$ is slowly varying along left-moving characteristics (that is, the time scale of its variation is the FRW time $t$), and ${\bf p}_-$ is slowly varying along right-moving characteristics.  Thus we can approximate their product at nearby points by the local product where the two geodesics intersect,
\begin{equation}
{\bf p}_-(s,\tau) \cdot {\bf p}_+(s',\tau) = - \alpha(s',\tau - \delta) + O(s - s') \ .   \label{ppa}
\end{equation}
Then
\begin{eqnarray}
\partial_\tau ({\bf p}_+ (s,\tau) \cdot {\bf p}_+(s',\tau)) &=&
 \frac{\dot a}{a} \biggl( \alpha(s',\tau - \delta) +  \alpha(s,\tau + \delta) \nonumber\\
&&\hspace{-90pt} -  \alpha (s,\tau)\,{\bf p}_+(s,\tau)\cdot{\bf p}_+(s',\tau) - \alpha  (s',\tau)\,{\bf p}_+(s,\tau)\cdot{\bf p}_+(s',\tau) \biggr)  + O([s - s']^2)\ .
\end{eqnarray}
When we integrate over a scale of order the Hubble time, the $\delta$ shifts in the arguments have a negligible effect $O(\delta)$ and so we ignore them.  Defining
\begin{equation}
h_{++}(s,s', \tau) = 1 - {\bf p}_+ (s,\tau) \cdot {\bf p}_+(s',\tau)\ ,
\end{equation}
we have
\begin{equation}
\partial_\tau h_{++}(s,s', \tau) =
 - \frac{\dot a}{a} h_{++}(s,s', \tau) [ \alpha(s',\tau ) +  \alpha(s,\tau) + O([s - s']^2)  ] \ .
\end{equation}
Thus
\begin{equation}
h_{++}(s,s', \tau_1) = h_{++}(s,s', \tau_0) \exp\Biggl\{ - \nu' \int_{\tau_0}^{\tau_1} \frac{d\tau}{\tau}
[ \alpha(s',\tau ) +  \alpha(s,\tau) + O([s - s']^2) ]  \label{hppint}
 \Biggr\} \ .
\end{equation}

Averaging over an ensemble of segments, and integrating over many Hubble times (and therefore a rather large number of correlation times) the fluctuations in the exponent average out and we can replace $\alpha(s,\tau )$ with $\bar\alpha = 1 - 2\bar v^2$,
\begin{equation}
\langle h_{++}(s,s', \tau_1) \rangle \approx
\langle  h_{++}(s,s', \tau_0) \rangle (\tau_1/\tau_0)^{- 2 \nu' \bar\alpha}\ . \label{avexp}
\end{equation}
Note that in contrast to previous equations the approximation here is less controlled.  We do not at present have a good means to estimate the error.  It depends on the correlation between the small scale and large scale structure (the latter determines the distribution of $\alpha$), and so would require an extension of our methods.  We do expect that the error is numerically small; note that if we were to consider instead $\langle \ln h_{++}(s,s', \tau) \rangle$ then the product in Eq.~(\ref{hppint}) would become a sum and  the averaging would involve no approximation at all.
  
Averaging over a translationally invariant ensemble of solutions, we have
\begin{equation}
\langle h_{++}(\sigma - \sigma', \tau_1) \rangle \approx
(\tau_1/\tau_0)^{- 2 \nu' \bar\alpha}\langle h_{++}(\sigma - \sigma', \tau_0) \rangle \ .
\end{equation}
We have used the fact that to good approximation (again in the sense of Eq.~(\ref{avexp})), $\epsilon$ is only a function of time, and so we can choose $\sigma = s - \int d\tau/\epsilon$ and $\sigma-\sigma' = s - s'$.  Equivalently,
\begin{equation}
\langle h_{++}(\sigma - \sigma', \tau) \rangle \approx \frac{ f(\sigma - \sigma') }{\tau^{2 \nu' \bar\alpha} }\ . \label{soln}
\end{equation}

The ratio of the segment length to $d_H = (1+\nu') t$ is
\begin{equation}
\frac{l}{d_H} \propto \frac{a \epsilon (\sigma - \sigma')}{t} \propto \frac{\sigma - \sigma'} {\tau^{1 + 2 \nu'\bar v^2 }}\ .
\end{equation}
The logic of our earlier discussion is that we use simulations to determine the value of $h_{++}$ at $l/d_H$ somewhat less than one, and then evolve to smaller scales using the Nambu action. That is,
\begin{equation}
h_{++}(\sigma - \sigma', \tau) = h_0 \ \mbox{when}\  \sigma - \sigma' = x_0 \tau^{1 + 2 \nu'\bar v^2 }\ , \label{match}
\end{equation}
for some constants $x_0$ and $h_0$.  We assume scaling behavior near horizon scale, so that $h_0$ is independent of time.  Using this as an initial condition for the solution~(\ref{soln}) gives
\begin{equation}
\langle h_{++}(\sigma - \sigma', \tau) \rangle \approx h_0 \Biggl( \frac{\sigma - \sigma'}{x_0 \tau^{1 + 2 \nu'\bar v^2 }}\Biggr)^{2\chi} \approx A (l/t)^{2\chi}\ ,\quad \chi = \frac{\nu'\bar\alpha}{1 + 2 \nu'\bar v^2 }
= \frac{\nu\bar\alpha}{1 - \nu\bar\alpha}
\ . \label{hpp}
\end{equation}
In the last form we have expressed the correlator in terms of physical quantities, the segment length $l$ defined earlier and the FRW time $t$.\footnote{We have not yet needed to specify numerical normalizations for $\sigma$ and $\tau$, or equivalently for $\epsilon$ and $a$.  The value of $h_0$ depends on this choice, but the value of $A$ does not.}

Eq.~(\ref{hpp}) is our main result.  Equivalently (and using $\sigma$ parity),
\begin{equation}
\langle {\bf p}_+ (\sigma,\tau) \cdot {\bf p}_+(\sigma',\tau) \rangle
= \langle {\bf p}_- (\sigma,\tau) \cdot {\bf p}_-(\sigma',\tau) \rangle
\approx 1 - A (l/t)^{2\chi}
\ .
\label{pppp}
\end{equation}
In the radiation era $\chi_r \sim 0.10$ and in the matter era $\chi_m \sim 0.25$.  There can be no short distance structure in the correlator ${\bf p}_+ \cdot {\bf p}_-$, because the left- and right-moving segments begin far separated, and the order $\dot a/a$ interaction between them is too small to produce significant nonsmooth correlation.  Thus, from~(\ref{ppa}) we get
\begin{equation}
\langle {\bf p}_+ (\sigma,\tau) \cdot {\bf p}_-(\sigma',\tau) \rangle
= -\bar\alpha + O(\sigma - \sigma')
\ . \label{pppm}
\end{equation}

\subsection{Small fluctuation approximation}

Before interpreting these results, let us present the derivation in a slightly different way.  The exponent $\chi$ is positive, so for points close together the vectors $ {\bf p}_+ (\sigma,\tau)$ and ${\bf p}_+(\sigma',\tau) $ are nearly parallel.  Thus we can write the structure on a small segment as a large term that is constant along the segment and a small fluctuation:
\begin{equation}
{\bf p}_+ (\sigma,\tau) = {\bf P}_+ (\tau) + {\bf w}_+ (\tau,\sigma) - \frac{1}{2} {\bf P}_+ (\tau)
w_+^2  (\tau,\sigma) + \ldots\ ,  \label{smallf}
\end{equation}
with $P_+(\tau)^2 = 1$ and ${\bf P}_+ (\tau) \cdot {\bf w_+} (\tau,\sigma) = 0$.  Inserting this into the equation of motion~(\ref{EOM2}) and expanding in powers of ${\bf w}_+$ gives
\begin{eqnarray}
\dot{\bf P}_+ &=& - \frac{\dot{a}}{a} \left[ {\bf P}_- - ({\bf P}_+ \cdot {\bf P}_-)\, {\bf P}_+ \right]\ ,
\\[6pt]
\dot{\bf w}_+ - \frac{1}{\epsilon} {\bf w}'_+ &=& \frac{\dot{a}}{a} \left[ ({\bf w}_+ \cdot {\bf P}_-)\, {\bf P}_+ + ({\bf P}_+ \cdot {\bf P}_-)\, {\bf w}_+ \right] \nonumber\\
&=& - ({\bf w}_+ \cdot \dot{\bf P}_+)\, {\bf P}_+ +  \frac{\dot{a}}{a} ({\bf P}_+ \cdot {\bf P}_-)\, {\bf w}_+\ .
\label{EOM3}
\end{eqnarray}
Since the right-moving ${\bf p}_-$ is essentially constant during the period when it crosses the short left-moving segment, we have replaced it in the first line of (\ref{EOM3}) with a $\sigma$-independent ${\bf P}_-$.  In the second line of (\ref{EOM3}) we have used ${\bf w}_+ \cdot {\bf P}_+ = 0$.  In the final equation for ${\bf w}_+$, the first term is simply a precession: ${\bf P}_+$ rotates around an axis perpendicular to both ${\bf P}_+$ and ${\bf P}_-$, and this term implies an equal rotation of ${\bf w}_+$ so as to keep ${\bf w}_+$ perpendicular to ${\bf P}_+$.  Eq.~(\ref{EOM3}) then implies that in a coordinate system that rotates with ${\bf P}_+$,
 ${\bf w}_+$ is simply proportional to $a^{-\alpha}$.

It follows that
\begin{equation}
\left\langle {\bf p}_+ (\sigma,\tau) \cdot {\bf p}_+(\sigma',\tau) \right\rangle - 1 = - \frac{1}{2} \left\langle [{\bf w}_+ (\sigma,\tau) - {\bf w}_+(\sigma',\tau)]^2 \right\rangle
\label{upup}
\end{equation}
scales as $a^{-2\bar\alpha}$ as found above (again we are approximating as in eq.~(\ref{avexp}), and again this statement would be exact if we instead took the average of the logarithm).  Similarly the four-point function of ${\bf w}_+$ scales as $a^{-4\bar\alpha}$.  We have not assumed that the field ${\bf w}_+$ is gaussian; the
$n$-point functions, just like the two-point function, can be matched to simulations near the horizon scale.  We can anticipate some degree of nongaussianity due to the kinked structure; we will discuss this further in section~3.2.

\subsection{Discussion}

Now let us discuss our results for the two-point functions. We can also write them as
\begin{eqnarray}
{\rm corr}_x(l,t) \equiv
\frac{ \left<{\bf x}'(\sigma,\tau)\cdot{\bf x}'(\sigma',\tau)\right> }{\left<{\bf x}'(\sigma,\tau)\cdot{\bf x}'(\sigma,\tau)\right>}  &\approx& 1 - \frac{A}{2(1 - \bar v^2)} (l/t)^{2\chi}\ ,
\nonumber\\
{\rm corr}_t(l,t) \equiv
\frac{\left< \dot{\bf x}(\sigma,\tau)\cdot \dot{\bf x}(\sigma',\tau)\right>}
{\left< \dot{\bf x}(\sigma,\tau)\cdot \dot{\bf x}(\sigma,\tau)\right>}  & \approx&
1 - \frac{A}{2 \bar v^2} (l/t)^{2\chi}\ . \label{corrs}
\end{eqnarray}
These are determined up to two parameters $\bar v^2$ and $A$ that must be obtained from simulations.
A first observation is that these scale, they are functions only of the ratio of $l$ to the horizon scale.  This is simply a consequence of our assumptions that the horizon scale structure scales and that stretching is the only relevant effect at shorter scales.  We emphasize that these results are for segments on long strings; we will discuss loops in Sec.~4.

It is natural to characterize the distribution of long strings in terms of a fractal dimension.
The RMS spatial distance between two points separated by coordinate distance $\sigma$ is
\begin{eqnarray}
\langle r^2(l,t) \rangle &=& \left<{\bf x}' \cdot{\bf x}' \right>  \int_0^l dl'  \int_0^l dl''\,
{\rm corr}_x(l' - l'',t) \nonumber\\
&\approx& (1 - \bar v^2) l^2 \left[ 1 -
\frac{A (l/t)^{2\chi}}{(2\chi+1)(2\chi+2)(1 - \bar v^2)} \right]\ .
\label{r2}
\end{eqnarray}
We can then define the fractal dimension $d_f$ (which is 1 for a straight line and 2 for a random walk),
\begin{eqnarray}
d_f &=& \frac{2 d \ln l}{d\ln \langle r^2(l,t) \rangle} \nonumber\\
&\approx& 1 + \frac{A \chi (l/t)^{2\chi}}{(2\chi+1)(2\chi+2)(1 - \bar v^2)} + O((l/t)^{4\chi})\ .
\label{fracdim}
\end{eqnarray}
The fractal dimension approaches 1 at small scales: the strings are rather smooth.
There is a nontrivial scaling property, not in the fractal dimension but rather in the deviation of the string from straightness,
\begin{equation}
1 - {\rm corr}_x(l,t) \propto (l/t)^{2\chi}\ , \quad
1 - {\rm corr}_t(l,t) \propto (l/t)^{2\chi}\ .  \label{crit}
\end{equation}
We define the scaling dimension $d_s = 2\chi$.  Note that $d_s$ is not large, roughly $0.2$ in the radiation era and $0.5$ in the matter era, so the approach to smoothness is rather slow.

\begin{figure}
\center \includegraphics[width=25pc]{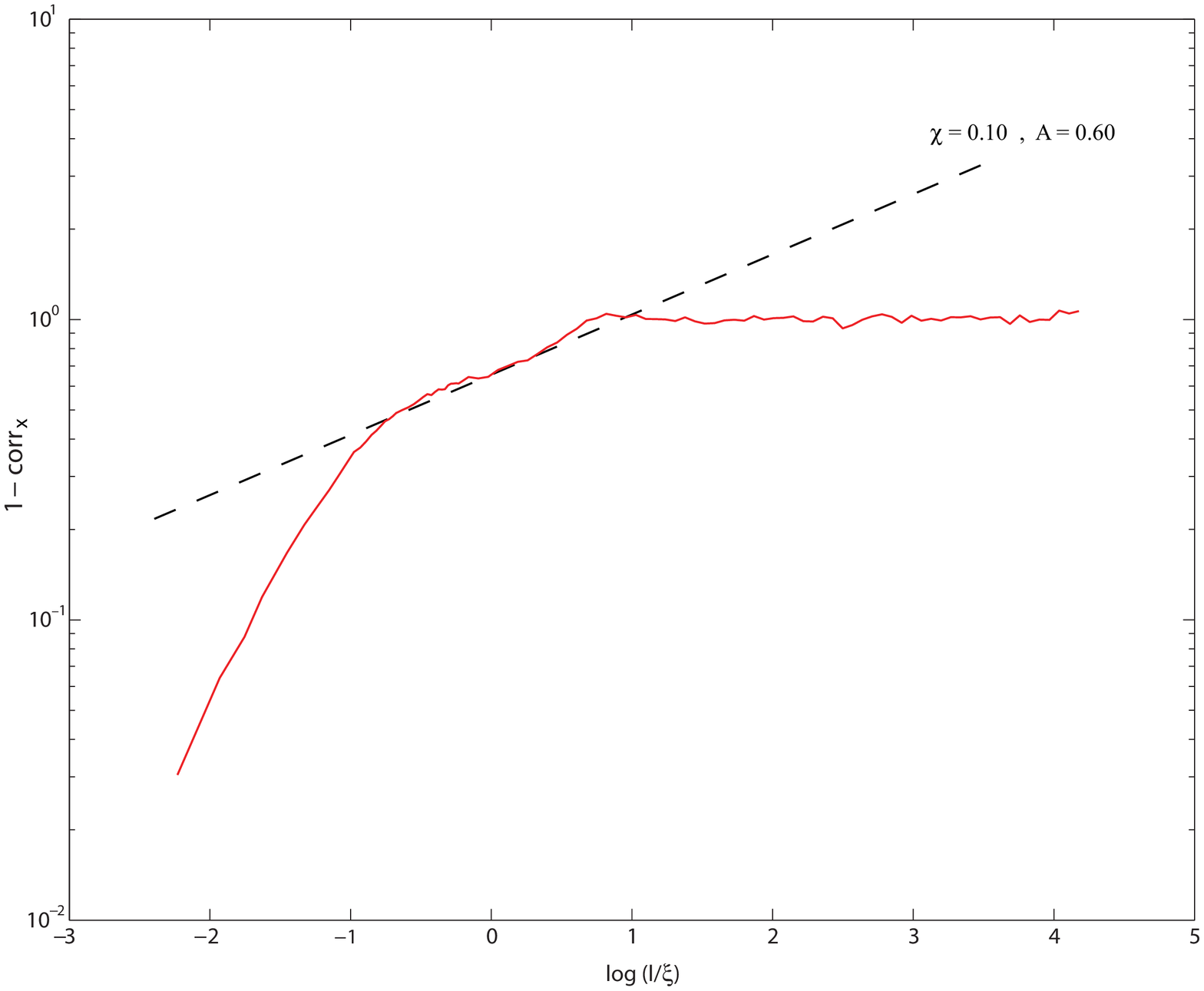}
\caption[]{Comparison of the model (dashed line) with the data provided by~\cite{MarShell2005} (solid red line) in the radiation-dominated era, for which the correlation length is $\xi \simeq 0.30 t$. } \label{mvd rad}
\end{figure}

Our general conclusions are in agreement with the recent simulations of Ref.~\cite{MarShell2005}, in that the fractal dimension approaches 1 at short distance.  To make a more detailed comparison it is useful to consider a log-log plot of $1 - {\rm corr}_x(l,t)$ versus $l$, as suggested by the scaling behavior~(\ref{crit}); we thank C. Martins for replotting the results of Ref.~\cite{MarShell2005} in this form.  The comparison is interesting.  At scales larger than $d_{\rm H}$ (which is $\sim 6.7 \xi$ in the radiation era and $\sim 4.3 \xi$ in the matter era)
the correlation goes to zero.  Rather abruptly below $d_{\rm H}$ the slope changes and agrees reasonably well with our result.  It is surprising to find agreement at such long scales where our approximations do not seem very precise.  On the other hand, at shorter scales where our result should become more accurate, the model and the simulations diverge; this is especially clear at the shortest scales in the radiation-dominated era (Fig.~\ref{mvd rad}).

\begin{figure}
\center \includegraphics[width=25pc]{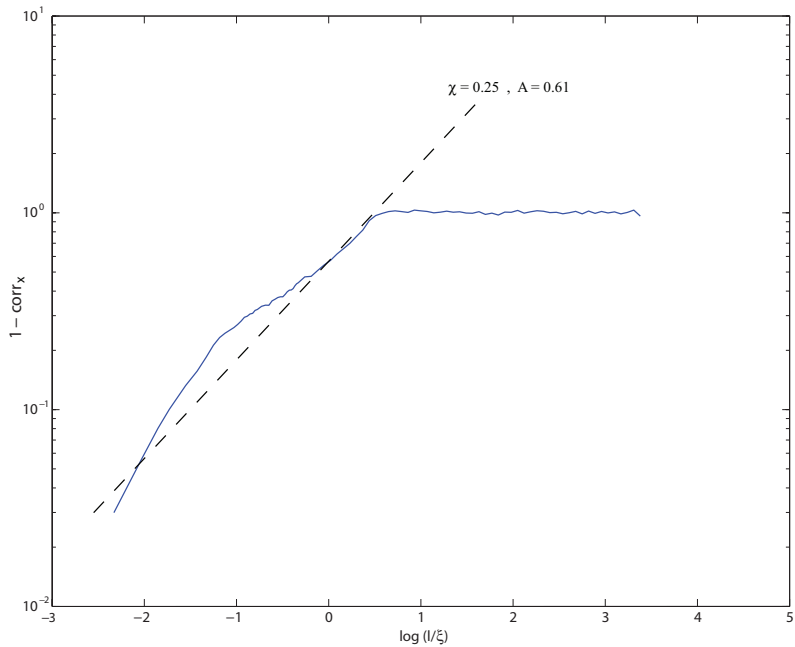}
\caption[]{Comparison of the model (dashed line) with the data provided by~\cite{MarShell2005} (solid blue line) in the matter-dominated era, for which the correlation length is $\xi \simeq 0.69t$.} \label{mvd mat}
\end{figure}

One possible explanation for the discrepancy is transient behavior in the simulations.  We have argued that the structure on the string is formed at the horizon scale and `propagates' to smaller scales (in horizon units) as the universe expands.  In Ref.~\cite{MarShell2005} the horizon size increases by a factor of order 3, and so even if the horizon-scale structure forms essentially at once, the maximum length scale over which it can have propagated is $3^{1 - \zeta}$, less than half an order of magnitude.  At smaller scales, the small scale structure seen numerically would be almost entirely determined by the initial conditions.  On the other hand, the authors of Ref.~\cite{MarShell2005}  (private communication) argue that their result appears to be an attractor, independent of the initial conditions, and that loop production may be the dominant effect.
Motivated by this we will examine loop production Sec.~4.  Indeed, we will find that this is in some ways a large perturbation, but we are still unable to identify a mechanism that would produce the two-point function seen in the simulations.  This is an important issue to be resolved in future work.

Thus far we have discussed ${\rm corr}_x$.  Our result~(\ref{corrs}) implies a linear relation between ${\rm corr}_x$ and ${\rm corr}_t$.  In fact, this holds more generally from the argument that there is no short-distance correlation between $ {\bf p}_+$ and ${\bf p}_-$, Eq.~(\ref{pppm}):
\begin{eqnarray}
(1-\bar v^2 )  {\rm corr}_x(l,t) - \bar v^2  {\rm corr}_t(l,t) &=& -\frac{1}{2}
\Bigl\langle {\bf p}_+ (\sigma,\tau) \cdot {\bf p}_-(\sigma',\tau) +  {\bf p}_- (\sigma,\tau) \cdot {\bf p}_+(\sigma',\tau) \Bigr\rangle \nonumber\\
&\to& \bar\alpha\ .
\end{eqnarray}
Inspection of Fig.~2 of Ref.~\cite{MarShell2005} indicates that this relation holds rather well at all scales below $\xi$.

The small scale structure on strings is sometimes parameterized in terms of an effective tension~\cite{Vilenkin:1990mz,Carter:1990nb}.  For a segment of length $l$ the effective tension is given by
\begin{eqnarray}
\frac{ \mu_{\rm eff} }{ \mu }  =  \frac{\sqrt{1 - \overline{v}^2} l}{\langle r(l) \rangle } 
\approx 1 + \frac{A}{2(1 - \overline{v}^2)(2\chi+1)(2\chi+2)}\left( \frac{l}{t} \right)^{2\chi} \ ,
\end{eqnarray}
where we have made use of result~(\ref{r2}).  Note that this is strongly dependent on the scale $l$ of the coarse-graining.

In conclusion, let us emphasize the usefulness of the log-log plot of $1 - {\rm corr}_x$.  In a plot of the fractal dimension, all the curves would approach one at short distance, though at slightly different rates.  The difference is much more evident in Figs.~1 and~2, and gives a clear indication either of transient effects or of some physics omitted from the model.

\subsection{The matter-radiation transition}

In this subsection we will assume that our stretching model is actually valid down the scale where gravitational radiation sets in.  If loop production or other relatively rapid processes are actually determining the small scale structure then this subsection is moot. 

We have noted that at very short scales we see structure that actually emerged from the horizon dynamics during the radiation era.  Thus we should take the radiation-to-matter transition into account in our calculation of the small-scale structure.  At the time of equal matter and radiation densities,
\begin{equation}
1 - \langle {\bf p}_+ (\sigma,\tau_{\rm eq}) \cdot {\bf p}_+(\sigma',\tau_{\rm eq}) \rangle
\approx A_{\rm r} (l_{\rm eq}/t_{\rm eq})^{2\chi_{\rm r}}
\ , \label{teq}
\end{equation}
where $l_{\rm eq}$ is the length of the segment between $\sigma$ and $\sigma'$ at $t_{\rm eq}$.
Assuming that the transition from radiation-dominated to matter-dominated behavior is sharp (which is certainly an oversimplification), we evolve forward to today using the result~(\ref{soln}). The right-hand side of Eq.~(\ref{teq}) is then multiplied by a factor $(t/t_{\rm eq})^{-2\nu_{\rm m}\bar\alpha_{\rm m}}$.  In terms of the length today, $l = l_{\rm eq} (t/t_{\rm eq})^{\zeta_{\rm m}}$, we have
\begin{equation}
1 - \langle {\bf p}_+ (\sigma,\tau) \cdot {\bf p}_+(\sigma',\tau) \rangle
\approx A_{\rm r} (l/t)^{2\chi_{\rm r}} (t/t_{\rm eq})^{-2\zeta_{\rm m}
- 2\chi_{\rm r}\zeta_{\rm m} + 2\chi_r}
\ . \label{radmat}
\end{equation}
This expression applies to scales $l(t)$ that, evolved backward in time, reached the horizon scale $d_H$ before the transition occurred, i.e. at a time $t_*$ defined by $l(t_*) \sim d_H$ such that $t_* < t_{\rm eq}$.  For longer scales, which formed during the matter era (for which $t_* > t_{\rm eq}$), we have simply
\begin{equation}
1 - \langle {\bf p}_+ (\sigma,\tau) \cdot {\bf p}_+(\sigma',\tau) \rangle
\approx A_{\rm m} (l/t)^{2\chi_{\rm m}}
\ . \label{radmat2}
\end{equation}

\begin{figure}
\center \includegraphics[width=25pc]{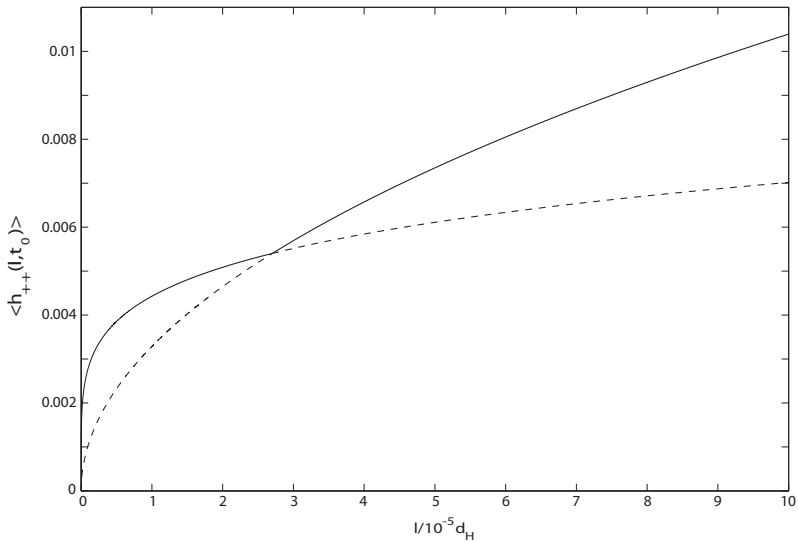}
\caption[]{Structure on the string, $\langle h_{++} \rangle$, as a function of the length $l$ at present time (solid curve).  On scales larger than the critical length $l_{\rm c} \sim 3\times 10^{-5} d_H$ the structure is determined by the matter era expression.  On scales below $l_{\rm c}$ the transition result~(\ref{radmat}) gives an enhanced effect.  The dashed curves show the extrapolations of the two relevant expressions: on small scales the actual structure is enhanced relative to the pure matter era result.} \label{transition}
\end{figure}

The transition between the two forms occurs along the curve determined by the intersection of the two surfaces~(\ref{radmat}) and~(\ref{radmat2}).  This determines the critical length at the time of equal matter and radiation densities, $l_{\rm c}(t_{\rm eq}) = (A_{\rm r}/A_{\rm m})^{1/(2\chi_{\rm m}-2\chi_{\rm r})}t_{\rm eq}$.  In terms of the length at some later time $t$, the transition occurs at
\begin{equation}
\frac{l_{\rm c}(t)}{t} \approx \left(\frac{A_{\rm r}}{A_{\rm m}}\right)^{1/(2\chi_{\rm m}-2\chi_{\rm r})} \left(\frac{t}{t_{\rm eq}}\right)^{\zeta_{\rm m}-1}
\ , \label{radmat3}
\end{equation}
so that the transition scale at the present time is $l_{\rm c} \sim 3\times 10^{-5} d_H$ (Fig.~\ref{transition}).  The transition result~(\ref{radmat}) implies more structure at the smallest scales than would be obtained from the matter era result, by a factor $(l_{\rm c}/l)^{2(\chi_{\rm m} - \chi_{\rm r})} \sim (l_{\rm c}/l)^{0.3}$.

Of course, precise studies of the small scale structure must include also the effect of the recent transition to vacuum domination; this period has been brief so the effect should be rather small.

\sect{Lensing}

Let us now consider the effect of the small scale structure on the images produced by a cosmic string lens.  Previous work has discussed possible dramatic effects~\cite{DeLaix1997,Bernardeau:2000xu}, including multiple images and large distortions.  We can anticipate that the rather smooth structure that we have found, which again we note is subject to our assumptions, will produce images with only small distortion.  We will use our stretching model of the two-point function.  If this proves incorrect one could apply the analysis using phenomenological values of $\chi$ and $A$; for example, the extrapolation of the results of 
Ref.~\cite{MarShell2005}  give a smoother string, and even less distorted images.

\subsection{Distortion of images}

\begin{figure}
\center \includegraphics[width=25pc]{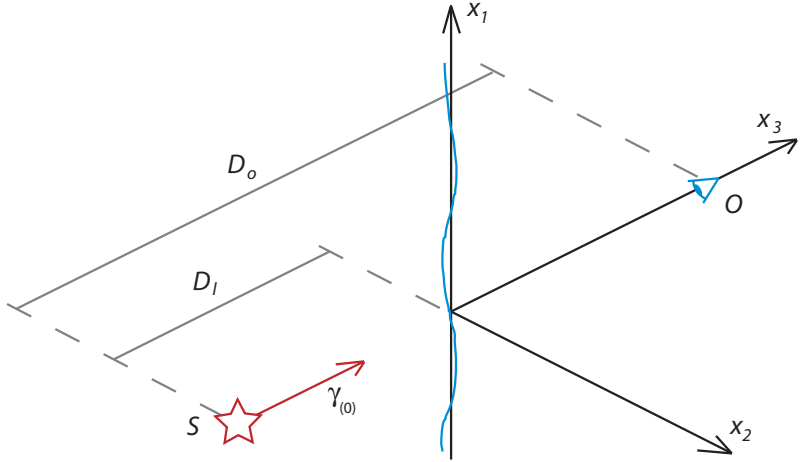}
\caption[]{Schematic representation of the system considered, with the string lens displayed along the $x_1$-axis and the distant source and the observer located at points $S$ and $O$ respectively.} \label{schematic}
\end{figure}

We quote the result of ref.~\cite{DeLaixVach1996} for the angular deflection of a light ray by a string,
\begin{equation}
\mbox{\boldmath $\gamma$}_\perp({\bf y}_{\perp}) = 4G\mu \int d\sigma \Biggl[ \frac{F_{\mu\nu} \gamma^\mu_{(0)} \gamma^\nu_{(0)}}{1 - \dot x_\parallel} \frac{ {\bf x}_\perp - {\bf y}_{\perp} }{
| {\bf x}_\perp - {\bf y}_{\perp} |^2 } \Biggr]_{t = t_0(\sigma)}\ .
\end{equation}
Here $\gamma_{(0)}^\mu$ is the four-velocity of the unperturbed light ray, which we take to be $(1,0,0,1)$ as shown in Fig.~\ref{schematic}, $y^\mu$ is a reference point on this ray, and subscripts $\perp$ and $\parallel$ are with respect to the spatial direction of the ray.  Also, $x^\mu(\sigma,t)$ is the string coordinate,\footnote{
The region where the the light ray passes the string is small on a cosmological scale, so to use our earlier results we can locally set $a = \epsilon = 1$, $dl = d\sigma$, $\partial_t = \partial_\tau$.}
in terms of which
\begin{equation}
F^{\mu\nu} = \dot x^\mu \dot x^\nu - x^{\mu\prime} x^{\nu\prime} - \frac{1}{2}\eta^{\mu\nu}
(\dot x^\rho \dot x_\rho - x^{\rho\prime} x_\rho^{\prime})\ ,
\end{equation}
and $t_0(\sigma)$ is defined by $t_0(\sigma) = x_3(\sigma,t_0) - y_3$.

In keeping with Sec.~2.3 we separate the string locally into a straight part and a fluctuation; we will keep the deflection only to first order in the fluctuation.  We consider here only the simplest geometry, in which the straight string is perpendicular to the light ray and at rest, so that ${\bf P}_+ = -{\bf P}_- = (1,0,0)$.  To first order in the fluctuation, $x^\mu(\sigma,t) = (t,\sigma,x_2(\sigma,t),x_3(\sigma,t))$.  One then finds
\begin{eqnarray}
\gamma_1({\bf y}_{\perp}) &=& 4 G \mu \int d\sigma\, \frac{- \dot x^{\vphantom2}_3(\sigma,t_0) (\sigma - y_1) + x_2^{\prime}(\sigma,t_0) y_2}{(\sigma - y_1)^2 + y_2^2} \ ,
\nonumber\\
\gamma_2({\bf y}_{\perp}) &=& - {\rm sgn}(y_2) \beta + 4 G \mu \int d\sigma\, \frac{\dot x^{\vphantom2}_3(\sigma,t_0) y_2 + x_2^{\prime}(\sigma,t_0) (\sigma - y_1)}{(\sigma - y_1)^2 + y_2^2}
\ ,
\end{eqnarray}
with $\beta = 4\pi G\mu$ being half the deficit angle of the string.
To the order that we work $t_0$ is a constant, corresponding to the time when the light ray is perpendicular to the straight string.

We can use our results for the small scale structure to calculate the two-point functions of the deflection.  Let us focus on the local magnifications parallel and perpendicular to the string, given by basic lensing theory as
\begin{eqnarray}
M_1 ({\bf y}_{\perp}) &=& 1 - \frac{D_l (D_o - D_l)}{D_o} \frac{\partial\gamma_1}{\partial y_1} ({\bf y}_{\perp})\ ,\nonumber\\
M_2 ({\bf y}_{\perp}) &=& 1 - \frac{D_l (D_o - D_l)}{D_o} \frac{\partial\gamma_2}{\partial y_2}
({\bf y}_{\perp})\ .
\end{eqnarray}
Here $D_o$ and $D_l$ are the distances of the source from the observer and the lens respectively (these would be the angular diameter distances on cosmological scales).  It is particularly interesting to consider the differential magnifications for the two images produced by a string,
\begin{equation}
\delta M_1 = M_1 ({\bf y}_{\perp}) - M_1 ({\bf y}'_{\perp})
\ ,\quad
\delta M_2 = M_2 ({\bf y}_{\perp}) - M_2 ({\bf y}'_{\perp})\ .
\end{equation}
We take for simplicity ${\bf y}_{\perp} = (0, b)$ and ${\bf y}'_{\perp} = (0, -b)$ for $b = \beta D_l (D_o - D_l)/D_o$; this corresponds to the symmetric images of an object directly behind the string.  Then
\begin{equation}
\delta M_1 = - \delta M_2 = -\frac{2b^2}{\pi} \int d\sigma\, x_2^{\prime}(\sigma,t_0) \partial_\sigma \left( \frac{1}{\sigma^2 + b^2} \right)\ .
\end{equation}

From Sec.~2 we obtain
\begin{eqnarray}
\left\langle {\bf w}_+ (\sigma,t) \cdot {\bf w}_+(\sigma',t) \right\rangle
&=& \left\langle {\bf w}_- (\sigma,t) \cdot {\bf w}_-(\sigma',t) \right\rangle
= 4  \left\langle x_2'(\sigma,t) x_2'(\sigma',t) \right\rangle
\nonumber\\
&=& A \Bigl\{  (\sigma/t)^{2\chi} + (\sigma'/t)^{2\chi} - ([\sigma - \sigma']/t)^{2\chi} \Bigr\}\ .
\label{u2p}
\end{eqnarray}
Note that Eq.~(\ref{upup}) does not fully determine the two-point functions~(\ref{u2p}), and in fact that the latter cannot be translation invariant.  We have fixed the ambiguity by defining the expectation value to vanish when $\sigma$ or $\sigma'$ vanishes (that is, at the point on the string nearest to the light ray); this amounts to a choice of how one splits ${\bf p}_\pm$ into ${\bf P}_\pm$ and ${\bf w}_\pm$.

As is stands, equation (\ref{u2p}) applies only when $|\sigma -\sigma'|$ is larger than the critical length $l_{\rm c}$.  If this is not the case, one must take into account the radiation-to-matter transition as discussed in section~2.5. According to equation (\ref{radmat}), the RHS of (\ref{u2p}) gets multiplied by a power of $t/t_{\rm eq}$.  From $\left\langle x_2' x_2' \right\rangle$ we obtain\footnote{The lensing scale, which is provided by $b$, is typically of the order of $10^{-7}d_H$, thus much smaller than the critical length $l_{\rm c}$.  We can then safely use the formula for $\left\langle x_2' x_2' \right\rangle$ valid for the smallest scales over the full range of integration. Corrections from the longest scales will increase the final result, but not significantly.}
\begin{equation}
\left\langle \delta M_1^2 \right\rangle
= \left\langle \delta M_2^2 \right\rangle
= \frac{\chi_{\rm r}(1-2\chi_{\rm r})}{2 \cos (\pi\chi_{\rm r})} A_{\rm r} (2b/t_0)^{2\chi_{\rm r}}(t_0/t_{\rm eq})^{-2\zeta_{\rm m}
- 2\chi_{\rm r}\zeta_{\rm m} + 2\chi_r}\ .
\end{equation}
Plugging in the numeric values for $\chi$, $A$ and $\zeta$ and using a representative value for the dimensionless parameter $G\mu \sim 10^{-7}$, we obtain a RMS differential magnification slightly below $1\%$:
\begin{equation}
\left\langle \delta M^2 \right\rangle^{1/2}
\simeq 0.009\ .
\end{equation}

\subsection{Alignment of lenses and nongaussianity}

Another question related to short-distance structure is the alignment of lenses.  Suppose we see a lens due to a long string, with a certain alignment.  Where should  we look for additional lens candidates?  Previous discussions~\cite{Huterer:2003ze,Oguri:2005dt} have considered the two extreme cases of a string that is nearly straight, and a string that is a random walk on short scales; clearly the networks that we are considering are very close to the first case.

We keep the frame of the previous section, with the lens at the origin in ${\bf x}_\perp$ and aligned along the 1-axis.  Then as we move along the string, the RMS transverse deviation is
\begin{eqnarray}
\langle x_2^2(l) \rangle &=& \int_0^l d\sigma'  \int_0^l d\sigma''\, \langle x'_2(\sigma)
x'_2(\sigma') \rangle \nonumber\\
&=& \frac{A l^2}{4(\chi + 1)} (l/t)^{2\chi}\ .
\end{eqnarray}
The extension in the $x$ direction is just $l$, so the RMS angular deviation is
\begin{equation}
\delta\varphi \sim
\sqrt{A / 4(\chi + 1)} (l/t)^{\chi} \equiv {\overline{\delta\varphi}}\ .
\end{equation}
If we put in representative numbers, looking at an apparent separation on the scale of arc-minutes for a lens at a redshift of order $0.1$,  we obtain with the matter era parameters a deviation ${\delta\varphi} \sim 0.05$ radians.  That is, any additional lenses should be rather well aligned with the axis of the first.  If the string is tilted by an angle $\psi$ to the line of sight, then projection effects increase $\delta\theta$ and $\delta\varphi$ by a factor $1/\cos\psi$.  Of course, for lensing by a {\it loop}, the bending will be large at lengths comparable to the size of the loop.  

Lens alignment provides an interesting setting for discussing the nongaussianity of the structure on the string.  If the fluctuations of $x_2'$ were gaussian, then the probability of finding a second lens at an angle $\delta\varphi$ to the axis of the first would be proportional to $e^{-\delta\varphi^2/2\overline{\delta\varphi}^2}$, and therefore very small at large angles.  However, we have considered thus far a typical string segment, which undergoes only stretching.  There will be a small fraction of segments that contain a large kink, and one might expect that it is these that dominate the tail of the distribution of bending angles.

Let us work this out explicitly.  Consider a left-moving segment of coordinate length $\sigma$, and let $P(\sigma,\tau,k)\, dk$ be the probability that it contain a kink for which the discontinuity $|{\bf p}_+ - {\bf p}_+'|$ lies between $k$ and $k+dk$ ($0 < k < 2$).  There are two main contributions to the evolution of $P$.  Intercommutations introduce kinks at a rate that we assume to scale, so that it is proportional to the world-sheet volume in horizon units, $a^2 \epsilon \sigma \tau^{-2\nu'/\nu} d\tau$, and to some unknown function $g(k)$.  Also, the expansion of the universe straightens the kink, $k \propto a^{-\bar\alpha}$~\cite{BB1989}.  Then\footnote{Other effects are often included in the discussion of kink density, such as the removal of kinky regions by loop formation~\cite{AllenCald1990}, but these have a small effect.}
\begin{equation}
\frac{\partial P}{\partial \tau} = \tau^{2\nu' (1 - \bar v^2 - 1/\nu)}
 \sigma g(k) + \bar\alpha \frac{\dot a}{a} \frac{\partial}{\partial k}(kP)\ . \label{pevol}
\end{equation}
 We set $P$ to zero at the matching time $\tau_0$ defined by
 \begin{equation}
\sigma= x_0 \tau_0^{1 + 2 \nu'\bar v^2 }\ ,
\end{equation}
as in Eq.~(\ref{match}): earlier kinks are treated as part of the typical distribution, while $P$ identifies kinks that form later.  For simplicity we assume that $x_0$ is small enough that the probability of more than one kink can be neglected.

To solve this, define
\begin{equation}
\kappa = k \tau^{\zeta'}\ , \quad
Q(\sigma,\tau,\kappa) = k P(\sigma,\tau,\kappa)\ .
\end{equation}
Then
\begin{equation}
\tau \partial_\tau Q = \sigma g(\kappa \tau^{-\zeta'}) \kappa \tau^{-1-\nu'}\ .
\end{equation}
This can now be integrated to give
\begin{eqnarray}
P(\sigma,\tau,\kappa) &=&
\sigma \tau^{\zeta'} \int_{\tau_0}^\tau \frac{d\tau'}{\tau'} \tau'^{-1-\nu'}
g(k \tau^{\zeta'} /\tau'^{\zeta'}) \nonumber\\
&=& \frac{\tilde\sigma}{\zeta'} k^{-1/\zeta}
\int_k^{k_0} \frac{dk'}{k'} k'^{1/\zeta} g(k')\ . \label{psol}
\end{eqnarray}
Here $k_0 = k (x_0 /\tilde\sigma)^{\zeta'/(1+2\nu'\bar v^2)}$ and $\tilde\sigma = \sigma/\tau^{1 + 2 \nu'\bar v^2}$. Note that $\tilde\sigma$ is just a constant times $l/t$, so the probability distribution scales.  The source $g(k)$ vanishes by definition for $k > 2$, so $k_0 > 2$ is equivalent to $k_0 = 2$.

Rather than the angle $\delta\varphi$ between the axis of the first lens and the position of the second lens, it is slightly simpler to consider the angle $\delta\theta$ between the two axes.  In the small fluctuation approximation this is just $x_2'(\sigma)$, and so the RMS fluctuation is
\begin{equation}
\overline{\delta\theta} = \sqrt{A/2}(l/t)^\chi\ .
\end{equation}
In the gaussian approximation, the probability distribution is $e^{-\delta\theta^2/2\overline{\delta\theta}^2}$ and so is very small for large angles.  On the other hand, a large angle might also arise from a segment that happens to contain a single recent kink.  Treating the segment as straight on each side of the kink, the probability density is then precisely the function $P(\sigma,\tau,k)\, dk$ just obtained, with $k = \delta\theta$.  If we consider angles that are large compared to $\overline{\delta\theta}$ but still small compared to 1, the range of integration in the solution~(\ref{psol}) extends essentially to the full range $0$ to $2$ and so the integral gives a constant.  Then
\begin{equation}
P(\sigma,\tau,k)\, dk \propto k^{-1/\zeta} dk = \delta\theta^{-1/\zeta} d\delta\theta \ .
\end{equation}
Thus the tail of the distribution is not gaussian but a power law, dominated by segments with a `recent' kink.  One finds the same for the distribution of $\delta\varphi$.  It is notable however that the exponent in the distribution is rather large, roughly 4 in the matter era and 10 in the radiation era.  Thus the earlier conclusion that the string is rather straight still holds.  The sharp falloff of the distribution also suggests that a gaussian model might work reasonably well, and indeed we will employ this in the next section.

\sect{Loop formation}

The model presented thus far assumes that stretching is the dominant mechanism governing the evolution of the small scale structure.  Motivated by the discrepancy with Ref.~\cite{MarShell2005} (and at the suggestion of those authors), we now consider the production of small loops.  Given the smoothness of the strings on short distances as found above, one might have thought that loop production on small scales would be suppressed, but we will see that this is not the case.  We will ignore loop reconnection, based on the standard argument that this is rare for small loops~\cite{Bennett:1989yp}.

We will treat the stretching model as a leading approximation, and add in the loop production as a perturbation.  However, we will find that the loop production diverges at small scales, and so the problem becomes nonlinear and our analytic approach breaks down.  Ultimately we would expect that some sort of improved scaling argument can be used to follow the structure to still smaller scales, but this is beyond our current methods; it is necessary first to identify the relevant processes.

\subsection{The rate of loop production}

In this section we aim to determine the rate of loop production, which occurs when a string self-intersects.  That is, we want to compute the number of loops $d\cal N$ with invariant length between $l$ and $l+dl$ originating from a self-intersection at a string coordinate between $\sigma$ and $\sigma + d\sigma$ during a time interval $(t,t+dt)$.  Loops much smaller than the horizon size evolve in an essentially flat space so we are able to use null coordinates $u = t + \sigma$ and $v = t - \sigma$.  In this case the left- and right-moving vectors ${\bf p}_\pm$ depend only on $u$ or $v$ respectively and the condition for the formation of a loop of length $l$ is ${\bf L}_+ = {\bf L}_-$ with
\begin{eqnarray}
{\bf L}_+(u,l) &\equiv& \int_u^{u+l} du'\, {\bf p}_+(u') \ , \nonumber \\
{\bf L}_-(v,l) &\equiv& \int_{v-l}^v dv'\, {\bf p}_-(v') \ .
\label{loopcondition}
\end{eqnarray}
Hence the number of loops formed is given by
\begin{equation}
d{\cal N} = \delta^{(3)}({\bf L}_+(u,l) - {\bf L}_-(v,l)) \left| \det {\bf J} \right| du\, dv\, dl \ ,
\label{dN}
\end{equation}
where ${\bf J}$ is the Jacobian for the transformation $(u,v,l) \rightarrow {\bf L}_+ - {\bf L}_-$.  This formalism is as in Refs.~\cite{Embacher:1992zk,ACK1993}.

Expanding around the stretching result in the form of section~2.3, we have the functional probability distribution
\begin{equation}
{\cal P}[{\bf p}_+, {\bf p}_-] \approx {\cal P}_0({\bf P}_+, {\bf P}_-)
{\cal P}_+[{\bf w}_+] {\cal P}_-[{\bf w}_-]\ .
\end{equation}
That is, with stretching alone the left- and right-movers cannot be correlated on small scales.  We will further assume that ${\cal P}_\pm$ are gaussian with variance given by the two-point function.  As discussed in Sec.~3.1 this two-point function is not completely determined by Eq.~(\ref{upup}), and in general we have
\begin{eqnarray}
\left\langle {\bf w}_+ (u') \cdot {\bf w}_+(u'') \right\rangle
&=& - A |u' - u''|^{2\chi}/t^{2\chi} + f(u') + f(u'') \nonumber\\
&\equiv& 2G(u' ,u'') \ .
\label{wpwp}
\end{eqnarray}
The function $G$ is defined for later reference, with the factor of 2 representing the sum over transverse directions. To specify the function $f$ one needs to choose how to split ${\bf p}_\pm$ into ${\bf P}_\pm$ and ${\bf w}_\pm$.  The natural choice here is to make ${\bf P}_\pm$ parallel to ${\bf L}_\pm$, i.e.\ ${\bf P}_\pm = {\bf L}_\pm/L_{\pm} $.  The transverse fluctuations along the segment that forms the loop, say $0 < \sigma < l$, then have vanishing average,
\begin{equation}
\int_u^{u+l} du'\, {\bf w}_+(u') = \int_{v-l}^v dv'\, {\bf w}_-(v') = {\bf 0} \ . \label{vanav}
\end{equation}
Integrating Eq.~(\ref{wpwp}) $\int_u^{u+l} du'$ and using the condition~(\ref{vanav}) to set this to zero gives $f(u') = \phi(u'-u)$ where 
\begin{equation}
\phi(\sigma) = \frac{A}{(2\chi+1) l t^{2\chi}} \left[ \sigma^{2\chi+1} + (l-\sigma)^{2\chi+1} - \frac{l^{2\chi+1}}{2\chi+2} \right]  \ .
\label{fsigma}
\end{equation}
Similarly, $\left\langle {\bf w}_- (v') \cdot {\bf w}_-(v'') \right\rangle = 2G(v',v'')$ with $f(v') = \phi(v-v')$.

The three-dimensional $\delta$-function in Eq.~(\ref{dN}) can be expressed in spherical coordinates as
\begin{equation}
\delta^{(3)}({\bf L}_+ - {\bf L}_-)  = \frac{1}{L_+^2} \delta(L_+ - L_-)\, \delta^{(2)}({\bf P}_+ -{\bf P}_-)\ . 
\label{deltas}
\end{equation}
We take the $z$-direction to be aligned along ${\bf P}_+ = {\bf P}_-$, 
in which case the Jacobian matrix is given by
\begin{equation}
{\bf J} = \left(
\begin{array}[tcb]{ccc}
w^x_+(l) - w^x_+(0)  & \; w^y_+(l) - w^y_+(0)  \; & \frac{1}{2}\!\left[ w^2_+(0) - w^2_+(l) \right] \\ \\
w^x_-(0) - w^x_-(l) & \; w^y_-(0) - w^y_-(l) \; & \frac{1}{2}\!\left[ w^2_-(l) -w^2_-(0) \right] \\ \\
w^x_+(l) - w^x_-(0) & \; w^y_+(l) - w^y_-(0) \; & \frac{1}{2}\!\left[ w^2_-(0) -w^2_+(l) \right]
\end{array}
          \right)  \ .
\end{equation}
For notational simplicity we have, after differentiating, set $u=0$ and $v=l$.
The fluctuations ${\bf w}_\pm$ effectively live in the $x$-$y$ plane due to the orthogonality condition ${\bf P}_\pm \cdot {\bf w}_\pm = 0$.

Let us first estimate the scaling of $d{\cal N}$ with $l$.  We have 
\begin{equation}
L_+ - L_- = \frac{1}{2} \int_0^l  d\sigma \left(w_+^2(\sigma) - w_-^2(\sigma) \right)  = 
O(l^{2\chi + 1} / t^{2\chi})\ . \label{lpm}
\end{equation}
The width of the distribution of $L_+ - L_-$ is of order $l^{1+2\chi}$ and so the average of $\delta(L_+ - L_-)$ is a density of order $l^{-1-2\chi}$.  The $\delta$-function~(\ref{deltas}) then scales as $l^{-3-2\chi}$ while the Jacobian scales as $l^{4\chi}$ (the columns are respectively of orders $l^{\chi}$, $l^{\chi}$, $l^{2\chi}$) , giving in all
\begin{equation}
d{\cal N} \propto \frac{dl}{l^{3 - 2\chi}}\ ,\quad
l \,d{\cal N} \propto \frac{dl}{l^{2 - 2\chi}}\ .
\label{dNscale}
\end{equation}
Thus, for $2\chi \leq 1$ (as found in both the matter and radiation eras), the total rate of string loss diverges at small $l$: the calculation breaks down.

We will discuss this cutoff further in the next subsection, but first we estimate the numerical coefficient.  We see no reason for a strong correlation between the $\delta$-function and Jacobian factors in $d{\cal N}$, so we take the product of their averages.  For the radial part of the $\delta$-function, a gaussian average gives
\begin{eqnarray}
\left< \delta(L_+ - L_-) \, \right> &=&  \int_{-\infty}^\infty \frac{dy}{2\pi} \, 
\left< e^{iy(L_+ - L_-)}\, \right> \nonumber\\
&=& \int_{-\infty}^\infty \frac{dy}{2\pi} \, 
e^{ iy \left< (L_+ - L_-) \right>_c - \frac{y^2}{2} \left< (L_+ - L_-)^2 \right>_c
- i \frac{y^3}{6} \left< (L_+ - L_-)^3 \right>_c + \ldots }
\nonumber\\
&=& \int_{-\infty}^\infty \frac{dy}{2\pi} \, 
e^{-y^2 R(\chi) + O(y^3) } \nonumber\\
&\approx& \frac{1}{\sqrt{4\pi R(\chi)}} \ .
\end{eqnarray}
The subscripts $c$ in the second line refer to the connected expectation values, obtained by contracting the gaussian fields ${\bf w}_\pm$ with the propagator~(\ref{wpwp}).  We have defined
\begin{eqnarray}
R(\chi) &=& \int_0^l d\sigma
\int_0^l d\sigma'\,G^2(\sigma,\sigma',t)  = \frac{l^{2 + 4\chi}}{t^{4\chi}} A^2 C(\chi)\ ,
\\
C(\chi) &=& \textstyle \frac{1}{4 (1 + 2\chi)^2}
\left({\frac{1 + 2\chi}{1 + 4\chi} + \frac{1}{(1 + \chi)^2} - \frac{4}{3 + 4\chi} - \frac{ 4 \Gamma^2(2 + 2\chi)}{\Gamma(4 + 4\chi)} } \right)\ .
\end{eqnarray}
Numerically $C(\chi) = (0.0049, 0.0106, 0.0111)$ for $\chi = (0.1, 0.25, 0.5)$.

We now turn to the angular part of the probability distribution function.  This must be a function of ${\bf P}_+ \cdot {\bf P}_-$, and we will take it to have exponential form~\cite{ACK1993}:
\begin{equation}
{\cal P}({\bf P}_+ , {\bf P}_-) = \frac{\lambda}{(4\pi)^2 \sinh \lambda} \, e^{-\lambda {\bf P}_+ \cdot {\bf P}_-}  \ .
\label{expform}
\end{equation}
The prefactor normalizes the distribution.  The parameter $\lambda$ can be determined from the requirement that $ \left< \, {\bf P}_+ \cdot {\bf P}_- \right>  = - \overline{\alpha} $ , or equivalently
\begin{equation}
\frac{1}{\lambda} - \frac{\cosh \lambda}{\sinh \lambda} \simeq - \overline{\alpha}  \ .
\end{equation}
Thus, $\lambda \sim 3 \overline{\alpha}$ .  More precisely, $\lambda$ is $0.55$ in the radiation epoch and $0.95$ in the matter epoch.

We are now able to calculate the average value of (\ref{deltas}).  Recognizing that $ \left< L_+^2 \right> \simeq l^2 $ we have
\begin{eqnarray}
\left< \delta^{(3)}({\bf L}_+ - {\bf L}_-) \right>  &\approx&
\frac{\lambda \left< \delta(L_+ - L_-) \, \right>}{(4\pi l)^2 \sinh \lambda}
\int d^2 {\bf P}_+\, d^2{\bf P}_-\, e^{-\lambda {\bf P}_+ \cdot {\bf P}_-} \delta^{(2)}({\bf P}_+ -{\bf P}_-) \nonumber \\
&\approx& \frac{\lambda \, e^{-\lambda} }{4\pi A \sqrt{4\pi C(\chi)} \sinh \lambda \, }\frac{t^{2\chi}}{ l^{3 +  2\chi}}  \ .
\label{avdelta}
\end{eqnarray}

We now consider $\left< |\det {\bf J}| \right>$.  In the gaussian approximation the relevant probability distribution is
\begin{equation}
{\cal P}_\pm({\bf w}_\pm(l) , {\bf w}_\pm(0)) = \frac{\det {\bf M}_\pm}{(2\pi)^2} \,
\exp \left[ -\frac{1}{2} ({\bf V}_\pm^i)^T {\bf M}_\pm {\bf V}^i_\pm \right] \ ,
\end{equation}
where the index $i$ is summed over the two coordinates $x$ and $y$, and
\begin{equation}
{\bf V}^i_\pm \equiv \left( \begin{array}{c} w^i_\pm(l) \\ w^i_\pm(0) \end{array} \right) \ .
\end{equation}
(That is, the columns and rows of the $2\times 2$ matrix ${\bf M}$ correspond to the points $0$ and $l$, not the index $i$.)
As usual, the whole distribution is determined solely by the two-point functions which have already been determined in~(\ref{wpwp}) and~(\ref{fsigma}),
${\bf M}_\pm^{-1} \delta^{ij}= \left< {\bf V}^i_\pm ({\bf V}^j_\pm)^T \right> $.  Thus, one finds that
\begin{equation}
{\bf M}_\pm = \frac{2(t/l)^{2\chi}}{A(1-\chi)}
\left( \begin{array}{cc} 1 & \chi \\ \chi & 1 \end{array} \right) 
\;\; , \;\;
\det {\bf M}_\pm = \left( \frac{2}{A} \right)^2 \left(\frac{t}{l}\right)^{4\chi} \frac{1+\chi}{1-\chi}
\ .
\end{equation}

We now have all we need to write down an expression for $\left< |\det {\bf J}| \right>$ .  Before we do so, let us perform a simplifying change of variables:
\begin{equation}
\begin{array}{lllllll} 
	X_\pm \equiv \sqrt{\frac{1+\chi}{1-\chi} \frac{1}{2A}} \left({t}/{l}\right)^\chi 
							 \left[ w^x_\pm(l) + w^x_\pm(0) \right]  \ ,     \\ [7pt]
	Y_\pm \equiv \sqrt{\frac{1}{2A}} \left({t}/{l}\right)^\chi
							 \left[ w^x_\pm(l) - w^x_\pm(0) \right]  \ ,	   \\ [7pt]
	Z_\pm \equiv \sqrt{\frac{1+\chi}{1-\chi} \frac{1}{2A}} \left({t}/{l}\right)^\chi
							 \left[ w^y_\pm(l) + w^y_\pm(0) \right]  \ ,     \\ [7pt]
	W_\pm \equiv \sqrt{\frac{1}{2A}} \left({t}/{l}\right)^\chi
							 \left[ w^y_\pm(l) - w^y_\pm(0) \right]  \ ,
				\end{array}
\end{equation}
under which the expectation value of the loop Jacobian takes a compact form:
\begin{equation}
\left< |\det {\bf J}| \right> =  \frac{A^2}{2\pi^4} \left(\frac{l}{t}\right)^{4\chi}
\int d^8{\bf X} \: e^{-{\bf X}^2} \: \left| F({\bf X}) + \frac{1-\chi}{1+\chi} \, G({\bf X}) \right|  \ .
\label{integral}
\end{equation}
Here we have defined an 8-dimensional vector ${\bf X}~\equiv~(X_+,X_-,Y_+,Y_-,Z_+,Z_-,W_+,W_-)$ and two functions of it:
\begin{eqnarray}
F({\bf X}) &\equiv& ( Y_+ W_- - Y_- W_+ ) \left( Y_-^2 + W_-^2 - Y_+^2 - W_+^2 \right)\ , \nonumber \\ 
G({\bf X}) &\equiv&  2 ( W_+ W_- - Y_+ Y_- ) (X_+ - X_-) (Z_+ - Z_-) +															\\
					 & & + \: (Y_+ W_- + Y_- W_+) \left[ (X_+ - X_-)^2 - (Z_+ - Z_-)^2 \right] \ .  \nonumber
\end{eqnarray}
Numerically the integral is $\simeq 110$ for the radiation epoch and $\simeq 94$ for the matter epoch; analytic estimates agree.

Finally, we can combine results~(\ref{avdelta}) and~(\ref{integral}).  Noticing that $du\, dv = 2 d\sigma\, dt$, we get
\begin{equation}
d \langle {\cal N} \rangle  =  \frac{c}{t^3} \left( \frac{l}{t} \right)^{2\chi-3} d\sigma \, dt \, dl
\label{loopdensity}
\end{equation}
with $c=0.121$ in the radiation epoch and $c=0.042$ in the matter epoch.

\subsection{The small loop divergence and fragmentation}

The total rate of string loss $\int l \, d \langle {\cal N} \rangle $ diverges at small $l$ for $\chi \leq 0.5$, as is the case in both the radiation and matter eras.  The fractal dimension approaches 1 at short distance, but the exponent $\chi$, which characterizes the approach to this limit, indicates a relatively large amount of short distance structure.  For example, if we consider the functions $p_{\pm}$ on the unit sphere, then $\delta \sigma \propto \delta p_{\pm}^{1/\chi}$: the effective fractal dimension is $1/\chi$, though the path is not continuous, there are gaps due to kinks.

Of course, the total rate of string loss is bounded, and so for small loops we must be multiply counting.  A precise account of this effect will be very complicated; see for example Ref.~\cite{ACK1993}, which parameterizes it in terms of a modification of the exponential distribution~(\ref{expform}) taking the form of a `hole' in the forward direction ${\bf p}_+ = {\bf p}_-$.   For the present we just determine the scale at which our calculation {\it must} break down.
In a scaling solution, the total amount of long string in a comoving volume scales as $\ell_\infty \propto a^3/t^2$, while stretching alone would give $\ell_\infty \propto a^{\bar\alpha}$.  The rate of string lost to loops must be
\begin{equation}
\frac{1}{\ell_\infty} \Biggl( \frac{\partial \ell_\infty}{\partial t} \Biggr)_{loops} = \frac{2 - 2 \nu(1 + \bar v^2)}{t}
\ .
\end{equation}
Our result for the same quantity, with a cutoff $l > l_*$, is
\begin{equation}
\frac{1}{\ell_\infty} \left( \frac{\partial \ell_\infty}{\partial t} \right)_{loops} = \int_{l_*}^\infty dl \, \frac{c l}{t^3} \left( \frac{l}{t} \right)^{2\chi-3}
= \frac{c}{(1-2\chi) t} \left( \frac{l_*}{t} \right)^{2\chi - 1} \ .
\end{equation}
Equating these gives
\begin{equation}
l_* \sim 0.18 t  \label{lstar}
\end{equation}
in both eras.

The cutoff~(\ref{lstar}) is rather large, just an order of magnitude below the horizon scale.  However, this is just the beginning of the story.  The formation of these large loops does not terminate the process of loop production on smaller scales; this continues on the large loop, unaffected at least initially by the separation of the loop from the infinite string.\footnote{By the same token, the fragmentation process begins even before the large loop pinches off.  By causality the reconnection that forms the loop cannot immediately have an effect on the fragmentation process in its interior.  Thus the large scale~(\ref{lstar}) may not be seen in the
primary loops as usually defined.  Rather, we should follow null rays from the two world-sheet points of primary loop formation backwards to the point where they meet, and define the length of the `causal primary' loop as twice the time difference so as to enclose the entire causally disconnected region.}
  Thus we must expect a rather dramatic fragmentation process.  The divergence of the rate of loop loss implies that at scales $l$ well below $l_*$ the na\"ive number density of loops of size $\sim l$ is greater than $1/l$: these become dense on the string.  Thus we might come to the conclusion that the fragmentation process continues until we end up with all loops at the gravitational radiation scale, the opposite of the na\"ive result~(\ref{lstar}) above.\footnote{The previous study~\cite{Scherrer:1989ha} may provide little guidance.  The small scale structure there was limited to 10 harmonics, and neither of the two spectra studied corresponds closely to our case: the less noisy was equivalent to $\chi = 0.5$, where the loop production is only logarithmically divergent, while the more noisy is dominated by harmonics near the cutoff and does not resemble our distribution.}

In fact, we believe that this is not precisely the situation either.  Rather, a non-self-intersecting loop will occasionally form, likely an order of magnitude or more smaller than the primary loop, and this length of loop will then be lost to the fragmentation process on smaller scales.  If a fraction $1/k$ of the string is lost in this way at each scale, then the total length of string in loops of size less than $l$ will scale as $l^{1/k}$.  The exponent may be rather small, so there could be a substantial production of small loops.  The divergent rate of loop production implies that our initial attempt to use analytic methods to separate scales has been too simple, but we might hope that a more sophisticated scaling model along the lines just indicated will be successful.

Let us briefly describe an analytic model of the formation of non-self-intersecting loops.  When the loop forms, we can define the distributions
\begin{equation}
\rho_\pm({\bf p}) = \frac{1}{l} \oint d\sigma\, \delta^2({\bf p} - {\bf p}_{\pm}(\sigma))\ .
\end{equation}
Because of the high fractal dimensions of the curves ${\bf p}_{\pm}(\sigma)$, we might think of these distributions as reasonably smooth functions.  The structure on the high harmonics of the loop will tend to smear these distributions into gaussians, but they will be skewed by the particular random values of the lowest harmonics.  Thus the distributions $\rho_+$ and $\rho_-$ will differ.  The production of the smallest loops is roughly local in ${\bf p}$, and so it can only continue until whichever of the two densities is smaller has been depleted.  This leaves residual distributions
\begin{equation}
\rho_+({\bf p}) - \rho_-({\bf p}) = +\rho^r_+({\bf p}) \ {\rm or}\  {-\rho^r_-}({\bf p})
\end{equation}
according to the sign of the difference at each point.  This nonvanishing difference implies a cutoff on the production of the smallest loops, and allows non-self-intersecting loops of finite size to form.

\subsection{Comparison with simulations}

The results that we have found have some notable agreements, and disagreements, with simulations.  One success is an apparent agreement with the recent simulations of Ref.~\cite{RSB2005} for the distribution of loop sizes: that reference finds a number density of loops per volume and per length $dn /dl \propto l^{-p}$
with $p = 2.5 \pm 0.1$ in the matter era and $p = 3.0 \pm 0.1$ in the radiation era.  We have exponents $3 - 2\chi = 2.5$ and 2.8 respectively.  Our exponent is for the production rate rather than the density, but in this regime the density is dominated by recently produced loops and so these are the same.  

Let us verify this, and also obtain the relative normalizations for the two quantities.  The number of loops per comoving volume of length between $l$ and $l+dl$ is unaffected by the expansion of the universe, and we are assuming that we are at scales where gravitational radiation can be neglected.  The number then changes only due to production:
\begin{equation}
\frac{d}{dt}\left( a^3 \frac{dn}{dl} \right) = \frac{a^3}{ \gamma^2 t^2} \frac{c}{l^3} \left( \frac{l}{t} \right)^{2\chi}\ ,
\end{equation}
where we have inserted the string scaling density $1/\gamma^2 t^2$.  This integral is dominated by recent times as long as $3\nu - 1 - 2\chi > 0$, as holds in both eras (again, $a \propto t^{\nu}$).  Integrating over time then gives
\begin{equation}
\frac{dn}{dl} = \frac{c}{(3\nu - 1 - 2\chi) \gamma^2 t^4} \left( \frac{l}{t} \right)^{-3 + 2\chi}\ .
\end{equation}
In the notation of Ref.~\cite{RSB2005},
\begin{equation}
C_\circ = \frac{c}{(3\nu - 1 - 2\chi) \gamma^2} (1 - \nu)^{-1 -2\chi}\ ,
\end{equation}
where the last factor is from the conversion from $d_H$ to $t$.  Using $\gamma =  0.59, 0.30$ respectively in the matter and radiation eras~\cite{MarShell2005}, this gives $C_\circ = 1.25$, 10.3.  These are larger than found in Ref.~\cite{RSB2005} by factors of 20 and 260 respectively.  
A large difference is not surprising, as we have argued that there will be substantial fragmentation which will move the loops to smaller $l$ where the distributions are larger.  However, it remains to be seen whether this can account for such an enormous difference. The fragmentation need not change the scaling as long as the long-string correlation functions remains a power law, because the shapes of the produced loops, and the resulting fragmentation, all scale; thus the agreement of the loop production exponents may be real.  

As we have noted, the loop scaling must eventually break down due to conservation of string, but with these reduced normalizations this happens at much lower scales, $\bar l_* \sim 4 \times 10^{-4}\, t$ and $2 \times 10^{-4}\, t$ respectively.  Below these scales we expect the distribution to fall as discussed in the fragmentation discussion.  
Thus we might expect the distribution to peak near these scales, but perhaps to be rather broad in both directions.  A possible phenomenological formula for loop production after fragmentation would be
\begin{equation}
\frac{ d {\cal N} }{d\sigma \, dt \, dl}  =  \frac{\bar c}{t^3} \frac{x^{2\chi-3} }
{1 + (x/x_*)^{2\chi -1 - 1/k} }\ , \quad x = l/t\ .
\end{equation}
The constant $\bar c$ is the reduced value of our $c$, determined by comparison with simulations such as ref.~\cite{RSB2005}, the exponent $1/k$, which is likely to be small, must be determined by improved simulations of fragmentation, and the value $x_* \sim \bar l_*/t$ is determined by conservation of total string.  Specific signatures, however, might be dominated by the tails, either at large loops or small.

The relatively large scale~(\ref{lstar}) is similar to that found in the recent paper~\cite{Vanchurin:2005pa}.  As discussed in Sec.~4.2, we expect that the final loop size is much smaller because of extensive fragmentation; the divergence that we have found for small loop formation will be equally present in the calculation of fragmentation of a large loop into smaller ones.  Ref.~\cite{Vanchurin:2005pa} finds a much less extensive fragmentation.  This points up the need for more complete studies of this process.  This should be possible within the context of flat spacetime simulations, beginning with long strings having the small scale structure as input (this does depend on resolving first the issue of the long string two-point function, to which we return below).

The production of very small loops takes place when ${\bf p_+} \sim {\bf p_-}$: that is, near cusps on the long strings as proposed in Refs.~\cite{Albrecht:1989mu,SiemOlum2003}.  Indeed, all the loops that we consider are rather smaller than the correlation length, so the functions ${\bf p_+}$ and ${\bf p_-}$ are each somewhat localized on the unit sphere, and necessarily in the same region since ${\bf L}_+ = {\bf L}_-$.  In this sense all of our loops are produced near cusps.  It is possible that there is a second population of loops that form at scales much longer than the correlation length, though most of these will quickly reconnect with long strings.  

The picture of a complicated fragmentation process is consistent with the results of Ref.~\cite{MarShell2005}.  Now let us return to the two-point functions as found there.  The partial agreement with our stretching model might now seem surprising, given the large production of small loops.  If the loop formation were distributed on the long strings, there would be at least one additional effect, where the string shortens due to loop emission and more distant (and therefore less correlated) segments are brought closer together.  However, the loop production is not distributed: it occurs when the functions ${\bf p}_{\pm}(\sigma)$, as they wander on the unit sphere, come close together.  Our picture is that whole segments of order $l_*$ (\ref{lstar}) are then excised.  Thus, most of the segments that remain on a long string at given time would have been little affected by loop production.

The discrepancy at short scales, which is particularly evident in the radiation era (Fig.~1), remains a puzzle.  Let us first note that at the shortest scales in Figs. 1 and 2 the slope of the curve approaches and possibly even exceeds unity, corresponding to the critical value $\chi = 0.5$: if the two-point function is of this form then the small-loop production converges.  Thus it is appealing to assume that it is feedback from production of short loops that accounts for the break in the curve in Fig.~1.\footnote{Note also that $\chi = 0.5$ corresponds to the functions ${\bf p}_\pm$ being random walks on the unit sphere.  This again suggests this value might be some dynamical fixed point.}
This is distinctly different from the cutoff via a ${\bf p}_+ - {\bf p }_-$  hole as discussed in Sec.4.2 and and in Ref.~\cite{ACK1993}: the latter is a modification of the $+-$ correlations (though one that would have little effect on the two-point function), whereas Fig.~1 shows a modification of the $++$ correlation.  

We are unable to identify a physical mechanism that would account for the observed two-point function.  Consider a very short segment on a long string.  With time, its size as a fraction of the horizon size decreases, so it moves to the left on Figs.~1 and~2.  If it experiences only stretching, it will follow the slope of the dashed line.  The loop production must have a strong bias toward removing segments with large fluctuations to account for the simulations.  However, the general picture above is that the initial loop formation occurs at the scale $l_*$, and so any small segments will be carried away without any bias on their own internal configuration.  Thus we would expect the two-point function to be given by the stretching power law down to arbitrarily small scales, until gravitational radiation enters.\footnote{One effect that may increase $\chi$ slightly is that segments that do not fall into the loop production `hole' at ${\bf p}_+ \sim {\bf p }_-$ may experience on average a larger value of $\alpha = -{\bf p}_+ \cdot {\bf p }_-$ and so have a reduced two-point  function --- see Eq.~(\ref{hppint}).  However, it seems impossible that this could account for the factor of 5 discrepancy of slopes evident in Fig.~1.}

We will not attempt to connect with the flat spacetime simulations of Refs.~\cite{Vanchurin:2005yb,Vanchurin:2005pa} because our approach requires expansion to drive the fractal dimension to 1 at short distance.

Finally we would like to note a significant puzzle regarding the loop velocities.  For a loop of length $l$, the mean velocity is
\begin{equation}
{\bf v} = \frac{1}{2l} ({\bf L}_+(u,l) + {\bf L}_-(v,l)) =  \frac{1}{l} {\bf L}_+(u,l) \ .
\end{equation}
Then
\begin{eqnarray}
\langle v^2(l) \rangle &=& \frac{1}{l^2} \int_0^l \!\!\!\int_0^l d\sigma\, d\sigma'\, \langle  {\bf p}_+ (\sigma) \cdot {\bf p}_+(\sigma') \rangle \nonumber\\
&=& 1 - \frac{2A}{(2\chi + 1)(2\chi + 2)} (l/t)^{2\chi}\ .
\end{eqnarray}
Looking for example at loops with $l = 10^{-2} t$, we obtain a typical velocity $0.985$ in the matter era and $0.90$ in the radiation era.  These are significantly larger than are generally expected; Ref.~\cite{Bennett:1989ch} gives values around $0.75$ and $0.81$ for loops of this size.  We emphasize that this is not a consequence of our dynamical assumptions, but can be seen directly by using the two-point functions of Ref.~\cite{MarShell2005} (the discrepancy is then even greater in the radiation era).  Assuming that both simulations are correct, we must conclude that the two-point function on loops is very different from that on long strings, in fact less correlated.  One possible explanation is that loop formation is biased in this way, but this seems unlikely: loops form when right- and left-moving segments of equal ${\bf L}(l)$ meet, and this will happen most often in the center of the distribution.  Rather, this seems to be another indication of the complicated nature of the fragmentation: the final non-self-intersecting loops must contain segments which began on the long string many times further separated than the final loop size.

\sect{Conclusions and future directions}

Our attempt to isolate the physics on different scales has not been completely successful, but it suggests some interesting directions for future work.  First of all, the two-point function on long strings must be better understood: is the discrepancy in Fig.~1 a result of omitted physics or numerical transients?  This question can be settled numerically, by studying sensitivity to initial conditions, and by studying the time-dependence of an ensemble of small segments.  A good understanding of the two-point function would then allow extrapolation to much smaller scales, for application to lensing (as in section~3), gravitational radiation~\cite{Vanchurin:2005pa}, and possible interference between short distance structure and cusps~\cite{SiemOlum2003}.

Second, our work points to correlations among the two-point function, loop production, fragmentation, and loop velocity.  In particular, our conjecture of a very complicated fragmentation process can be tested --- for example, the argument that non-self-intersecting loops must contain segments that originated far apart on the long string.  Also, the normalization difference between our loop production and that in Ref.~\cite{RSB2005} must presumably be understood in terms of fragmentation.

In conclusion, cosmic string networks are probably not as complicated as turbulence, but they share the property that a rather simple set of classical equations leads to a complicated dynamics that challenges both numerical and analytical attacks.  We hope that our work is a step toward a more unified understanding of the properties of these networks.

\section*{Acknowledgments}

We are grateful to Carlos Martins and Paul Shellard for providing the data from their numerical simulations and for valuable comments and suggestions.  We also thank Ed Copeland and Thibault Damour for discussions and Xavier Siemens for comments. 
JVR acknowledges financial support from {\it Funda\c{c}\~ao para a Ci\^encia e a Tecnologia}, Portugal, through grant SFRH/BD/12241/2003.  This work was supported in part by NSF grants PHY99-07949 and PHY04-56556.


\end{document}